\documentclass[prb,floAtfix,aps,epsf,twocolumn,showpacs]{revtex4-1}
\usepackage[dvips]{graphics}
\usepackage{graphicx}
\usepackage[utf8]{inputenc}
\usepackage{amssymb}
\usepackage{epstopdf}
\usepackage{bm}
\usepackage{dcolumn}
\usepackage{epsfig}
\usepackage{latexsym}
\usepackage{amsmath}
\usepackage{color}
\usepackage{array}
\usepackage{enumerate,dsfont}

\begin{document}
\title{Towards a realistic transport modeling in a superconducting nanowire with Majorana fermions}
\author{Diego Rainis, Luka Trifunovic, Jelena Klinovaja, and Daniel Loss}
\affiliation{Department of Physics, University of Basel, Klingelbergstrasse 82, CH-4056 Basel, Switzerland}
\date{\today}
\pacs{74.45.+c, 73.63.Nm, 74.78.Na}

\begin{abstract}
Motivated by recent experiments  searching for Majorana fermions (MFs)
in hybrid semiconducting-superconducting nanostructures, we consider a realistic tight-binding model and analyze its transport behavior numerically. 
In particular, we take into account the presence of a superconducting contact, used in real experiments to extract the current, which is usually not included in theoretical calculations.
We show that important features emerge that are absent in simpler models, such as the shift in energy of the proximity gap signal, and 
the enhanced visibility of the topological gap for increased spin-orbit interaction. 
We find oscillations of the zero bias peak as a function of the magnetic field and study them analytically.
We argue that many of the experimentally observed features hint at an actual spin-orbit interaction larger than the one typically assumed.
However, even taking into account all the known ingredients of the experiments and exploring many parameter regimes for MFs, we are not able to reach full agreement with the reported data. 
Thus, a different physical origin for the observed zero-bias peak cannot be excluded.

\end{abstract}

\maketitle

\section{Introduction}
\label{sec:Introduction}

The experimental search~\cite{mourik_signatures_2012,deng_observation_2012,das_evidence_2012} of Majorana fermions (MFs) predicted to occur in condensed matter systems~\cite{fujimoto_topological_2008,sato_non-abelian_2009,sau_generic_2010,sato_non-abelian_2010,alicea_majorana_2010,DasSarma_Majorana_2010,Oreg_Majorana_2010} is 
challenging due to the fact that MFs are characterized by
zero coupling to electromagnetic fields.  
Only an indirect identification is possible,
in particular via a zero-bias conductance peak (ZBP) ~\cite{Sengupta_organicSC_2001,ZBP-DasSarma}. 
However, such features are not an unambiguous demonstration of MFs. The same ZBPs can be induced 
by  different mechanisms, including the Kondo effect~\cite{Kondo_Sasaki}, Andreev bound states~\cite{ABS}, weak antilocalization and reflectionless tunneling~\cite{Reflectionless}.

A typical experimental setup~\cite{mourik_signatures_2012,deng_observation_2012,das_evidence_2012}  (see Fig.~\ref{fig:SchematicN-NW-SW-S}) consists of a semiconducting nanowire with Rashba spin-orbit interaction (SOI) deposited on or coated with a bulk $s$-wave superconductor (S) on one end and contacted through a tunnel barrier by a normal lead, on the other end. 
Part of the nanowire is  in a superconducting state induced by proximity effect. 

The transition to the topological phase controlled by a magnetic field $B$ is accompanied by a closing and reopening of the excitation 
gap~\cite{sato_non-abelian_2009,sau_generic_2010,sato_non-abelian_2010,alicea_majorana_2010,DasSarma_Majorana_2010,Oreg_Majorana_2010}. The topological phase  persists for all $B$-fields above a critical $B_{\rm c}$ 
in a one-band model, while it could have a finite upper critical field in a multiband model, where bands cross at large fields and hybridization of MFs  takes place. However, in experiments one typically explores regimes where only one band undergoes a  transition~\cite{mourik_signatures_2012,deng_observation_2012,das_evidence_2012}. 
For a topological section of finite length $L_\star$, the MFs at each end with localization length $\xi_{\rm M}$ depending on $B$  can overlap, leading to splitting of the ZBP at strong $B$-fields.

The experiments~\cite{mourik_signatures_2012,deng_observation_2012,das_evidence_2012} show features which are {\it partially} consistent with the existence of MFs.
However, {\it quantitative} agreement with the theory is still missing, and in particular the following points have to be clarified:

(i)~The most evident discrepancy between experiment and theory is the absence of any experimental signature  of the excitation gap in the nanowire. Recently, this fact has been ascribed to the spatial distribution of the wave functions for low chemical potential $\mu$~\cite{stanescu_close_2012,pientka_enhanced_2012}. 

(ii)~The ZBP in the experiments appears above a certain magnetic field, persists over a finite range of $B$, and then disappears, rather than splitting as expected for MFs.

(iii)~The ZBP conductance is not quantized, with values being much smaller than  $2e^2/h$~\cite{mourik_signatures_2012,deng_observation_2012,das_evidence_2012}, whereas   MFs are predicted to give  $2e^2/h$~\cite{Sengupta_organicSC_2001,Ng_ResonantAndreev_2009,Beenakker_QuantizedConductanceDisorder_2011}.

(iv)~The proximity-induced gap $\Delta_\star$  depends only weakly on $B$ in the $dI/dV$ curves, and the corresponding conductance  decreases significantly for large $B$~\cite{mourik_signatures_2012,das_evidence_2012}. Such a sudden reduction is not predicted, and the gap should close much faster for increasing $B$. 
This issue has not been pointed out in previous theoretical studies.

To address the above issues, we perform numerical calculations of the two-terminal conductance $G$ in a hybrid structure, referred to as NSS$^\prime$ setup, shown 
in Fig.~\ref{fig:SchematicN-NW-SW-S}, which closely models the experiment. 
Here, $G$ is calculated within the standard scattering theory~\cite{LambertRaimondi1998}, with the help of the recursive Green's function techniques~\cite{RecursiveGFMacKinnon}. 
This allows us to model a complex structure close to experiment that is not amenable to analytical approaches~\cite{Note}.

To be specific, we focus on InSb nanowires~\cite{mourik_signatures_2012,deng_observation_2012} and we use as a primary reference the experiment of Ref.~\cite{mourik_signatures_2012}, with exceptions  described below.
Nonidealities such as multiple occupied subbands, disorder, finite width of electrostatic barriers, finite coherence lengths, and nonzero temperature are taken into account. 

Our study reveals important features not emphasized so far. 
For this, the presence of the bulk superconductor turns out to be decisive.
We  summarize here our main findings. 
In our NSS$^\prime$ setup, the gap-edge conductance peak decreases in intensity for increasing $B$, a feature that is also not captured by simpler models.
Further, in some regimes the closing of the gap becomes visible in the conductance, while it does not in an NS setup.
We find oscillations of the ZBP as a function of $B$ and explain their origin.
We argue that disorder is unlikely to be the explanation of the observed ZBPs. Further, we show that the tunnel barrier plays an important role for the visibility of peaks.  
Finally, according to our results the experimental $dI/dV$ behavior seems to point to a SOI strength larger than the one reported.

\section{Model}
We consider a two-dimensional rectangular nanowire of
length $L$ along the $\hat {\bm x}$ direction and lateral extension $W$ in the
$\hat{\bm y}$ direction. All the plots presented in this manuscript refer to 4-subband wires ($W=4$), but we have conducted similar simulations for $W=1,2,8$ as well, noting only quantitative changes in the relative strength of the different $dI/dV$ features (besides the known peculiarity of the one-band case, where some features are absent).
\begin{figure}[h!]
\includegraphics[width=\columnwidth]{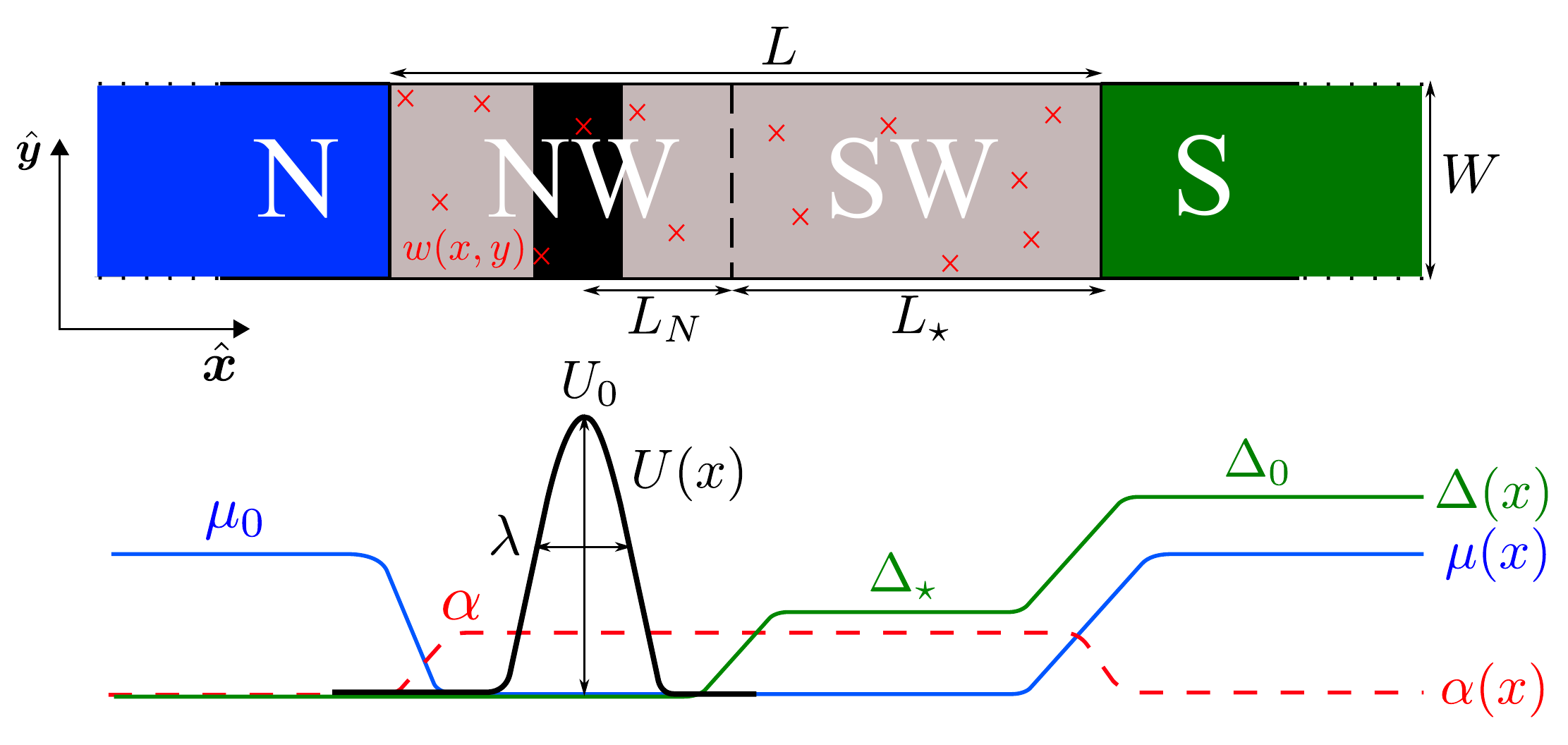}
\caption{The schematics of the NSS$^\prime$ geometry setup we consider in this
work (top panel). The nanowire (gray) is connected on the left to a
semi-infinite normal lead ({N}, blue) and on the right to a semi-infinite bulk
$s$-wave superconducting lead ({S}, green). It consists of a normal 
section (NW, gray), where a potential barrier $U(x)$ (black) is created, and a
proximity-induced superconducting nanowire section ({SW}, gray). 
We allow for static disorder $w(x,y)$ (red crosses) in the nanowire.
The spatial dependence of  the parameters entering the Hamiltonian in Eq.~(\ref{H}) is
qualitatively depicted in the bottom panel.} 
\label{fig:SchematicN-NW-SW-S}
\end{figure}

The tight-binding Hamiltonian (lattice constant $a$) describing the different sections of the setup 
has the form
\begin{align}
  \label{H}
  H = & \sum_{\bm m,\bm d} c^\dagger_{\bm m+\bm{ d},\alpha}
  \left[ -t \delta_{\alpha\beta}-i\bar\alpha_{\bm m}
  (\bm{\hat x}\cdot\bm{d})\sigma^y_{\alpha\beta} \right]c_{\bm{m},\beta}\nonumber\\
  +&\sum_{\bm m}c^\dagger_{\bm m,\alpha}\left[
  (\epsilon_{\bm m}-\mu_0)\delta_{\alpha\beta} - \frac{g_{\bm m}}{2}\mu_{\rm B}
  B_x\sigma^x_{\alpha\beta} \right]c_{\bm m,\beta} \nonumber \\
  +&\sum_{\bm m}\Delta_{\bm m} \left( c^\dagger_{\bm m,\uparrow}c^\dagger_{\bm
  m,\downarrow}+\text{H.c.} \right),
\end{align}
where $t=\hbar^2/(2ma^2)$ is the hopping amplitude (set to 1 and taken as an energy unit) and $\bar\alpha$ is the spin-flip hopping amplitude, related to the physical SOI parameter by $\bar\alpha=\alpha/2a$ and to the SOI energy by $E_{\rm so}=\bar\alpha^2/t$. 
Here and in the remainder of the paper we are neglecting transverse spin-orbit coupling, but we have checked that the introduction of a small finite transverse SOI is not affecting qualitatively our results.
We made the assignment $t=10$ meV, which corresponds to taking $a\simeq15$ nm, and realistic sizes ($\sim\mu$m) are then amenable to reasonable computations.
The sums run over all lattice sites $\bm m$ and  nearest neighbors $(\bm m+\bm d)$. Implicit summation over repeated spin indices is assumed. The constant $\mu_0$ is chosen to set the common chemical potential to the zero-field bottom of the topmost band and  depends on the number of  subbands (i.e. on $W$). 
Further, $\epsilon_{\bm m}=-\mu_{\bm m}+U_{\bm m}+w_{\bm m}$ accounts for local variations of the chemical potential, for the tunnel-barrier potential $U_{\bm m}$, and includes an on-site random potential $w_{\bm m}$ which models Anderson disorder. The tunnel barrier has a Gaussian profile with height $U_0$ and width $\lambda$. The external magnetic field $\bm B$ points along the nanowire axis ($ \hat {\bm x}$) and induces a Zeeman splitting $2V_{\rm Z}={g_{\bm m}\mu_{\rm B}}B$. 
Finally, $\Delta$ is the pairing amplitude and can either account for the native superconductivity in the bulk $s$-wave superconducting lead ($\Delta_0$) or for the proximity-induced pairing in the nanowire ($\Delta_\star$), as exemplified in Fig.~\ref{fig:SchematicN-NW-SW-S}. 
All the above quantities are taken to be site-dependent along the $\hat {\bm x}$ direction (except $w_{\bm m}$, which is taken to be completely random), so that we can model  different parts of the setup. 
The normal lead is characterized by 
\begin{align}\nonumber
\bar\alpha&=0, ~ \mu\simeq-\mu_0 ~{\rm (i.e.~metallic~regime)}, \\
g&=2, ~ w_{\bm m}=0, ~\Delta_{\bm m}=0.
\end{align}
The nanowire is characterized by finite $\bar\alpha=\bar\alpha_{\rm R}$, chemical potential $\mu\simeq0$ close to the bottom of the topmost band, $g=50$ appropriate for InSb nanowires, and $\Delta_{\bm m}$ varying from 0 in the normal section to $\Delta_\star$ in the proximized section.
The nanowire is adiabatically connected to a metallic superconducting lead with 
\begin{align}\nonumber
\bar\alpha&=0, ~\mu\simeq -\mu_0,\\
 g&=2, ~w_{\bm m}=0, ~\Delta_{\bm m}=\Delta_0\geq\Delta_\star. 
\end{align}

\begin{figure}[h!]
\includegraphics[width=1.\columnwidth]{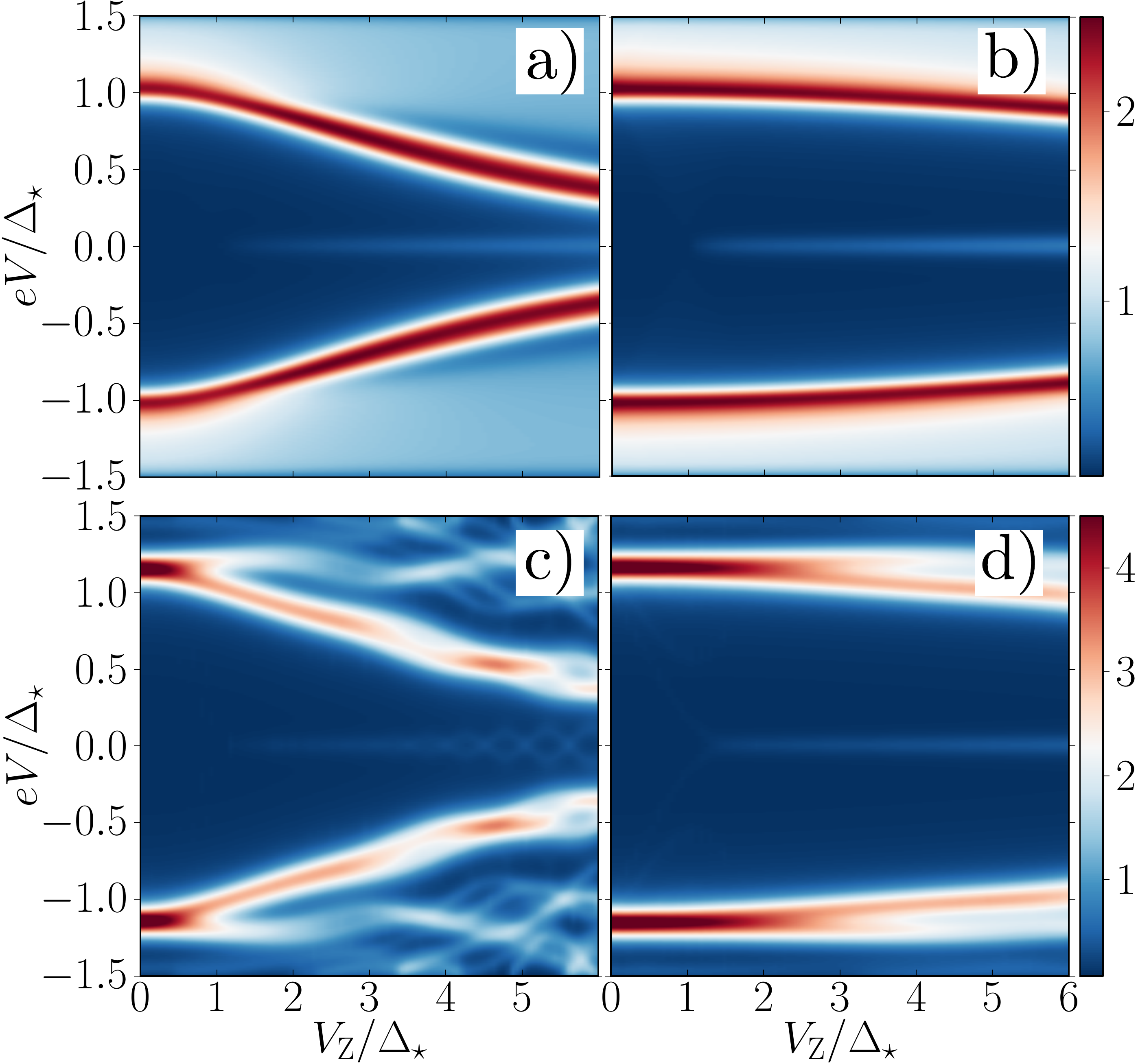}
\caption{Effect of larger SOI strength, clean case. We plot here the differential conductance $dI/dV$ evaluated as a function of bias voltage $V$ and Zeeman energy $V_{\rm Z}$. 
Panels (a) and (b) refer to the NS configuration, while (c) and (d) refer to the NSS$^\prime$ setup. 
The parameters used here correspond to: $\Delta_\star=250~\mu$eV, $\Delta_0=2.1$ meV (only NSS$^\prime$), $\mu_0=-3.8$ meV, $U_0=45$ meV, $\lambda=1$ nm (narrow barrier), $L_{\rm N}=0$, $L_\star=3~\mu$m (only NSS$^\prime$) and $\mu=0$, which corresponds to a bulk critical $V^{\rm c}_{\rm Z}=\Delta_\star$. For the case of InSb, the plotted range $V_{\rm Z}=0-6\Delta_\star$ corresponds to $B=0-1$ T. Temperature is set to $T=75$ mK.
$\alpha=0.2$ eV$\cdot$\AA (left column). $\alpha=0.8$ eV$\cdot$\AA (right column). 
Larger SOI yields a slower closing of the $k_{\rm F}$-gap $\Delta_{k_{\rm F}}(B)$, in both configurations, where $k_{\rm F}$ is the Fermi momentum. Notice that in the NSS$^\prime$ case the $k_{\rm F}$-gap signal decreases in intensity as the magnetic field is increased. }
\label{fig:Slow_closing_gap}
\end{figure}

In a simpler model the nanowire is semi-infinite, without external superconductor, referred to as NS geometry. This corresponds to taking the superconducting lead to be identical to the nanowire, with a single pairing amplitude $\Delta_\star$. In such a configuration, the second MF is always moved to infinity, and the ZBP is locked to zero for all  $B>B_{\rm c}$, whereby the topological transition occurs at the ``bulk'' critical field  $({g\mu_{\rm B}}/{2})B_{\rm c}=\sqrt{\Delta_\star^2+\mu^2}$~~\cite{sato_non-abelian_2009,sau_generic_2010,sato_non-abelian_2010,alicea_majorana_2010,DasSarma_Majorana_2010,Oreg_Majorana_2010}. 
We will sometimes switch to this NS configuration in order to connect with previous studies~\cite{PascalABS2012,prada_transport_2012,LinZBP2012,stanescu_close_2012,pientka_enhanced_2012,Altland_disorder_2012,Pikulin_disorder_2012,PascalCL2012} 
and to understand the effect of the bulk superconductor. 

In the actual experiments, and in a fully microscopic theoretical simulation, the nanowire has zero pairing everywhere, and the effective gap $\Delta_\star$ is generated by the coupling to the bulk superconductor. Usually one can forget about the superconductor and work with a wire with given $\Delta_\star$. However, in the considered setup the bulk S is still playing a role, since current is extracted through it, and it is therefore substantially modifying the $dI/dV$ behavior (not simply by singling out the Andreev reflection contribution of an NS calculation). It would be different in the case of transport across a proximity wire placed on a superconductor that is not used as a contact (NSN geometry).

Our setup aims exactly at taking this fact into account: The proximity effect is included in an effective fashion (not microscopically), but we do have two different pairing regions that electrons have to cross.
Still, with the sequential geometry of Fig.~\ref{fig:SchematicN-NW-SW-S} we are slightly simplifying here the  experimental setup~\cite{mourik_signatures_2012,deng_observation_2012,das_evidence_2012}, where the nanowire is side-contacted, or top-contacted, and the current does not follow a straight path.\\

First we note that the value of the SOI $\alpha$ in the experiments is not known, as also noticed in Ref.~\onlinecite{liu_zero-bias_2012}, since the only available measurements have been performed in a different setup, where the SOI was likely modified. Similarly, the proximity pairing amplitude is not directly accessible, and one can only deduce it from the $dI/dV$ behavior. Thus, it becomes interesting and even necessary to consider regimes with different SOI strengths, or different proximity pairing amplitudes.

\section{Discussion}
The first important point we want to make is that by assuming that the actual SOI is larger than the reported one (e.g., $\alpha=0.2$~eV$\cdot$\AA, or $E_{\rm so}=50~\mu$eV in Ref.~\onlinecite{mourik_signatures_2012}), one can get a substantial improvement in the calculated $dI/dV$ behavior, with features more similar to experiments~\cite{mourik_signatures_2012,deng_observation_2012,das_evidence_2012}. 
In other words, the measured data  suggest a stronger SOI.  
In particular, we observe the following facts.

\begin{figure}[h!]
\includegraphics[width=0.80\columnwidth]{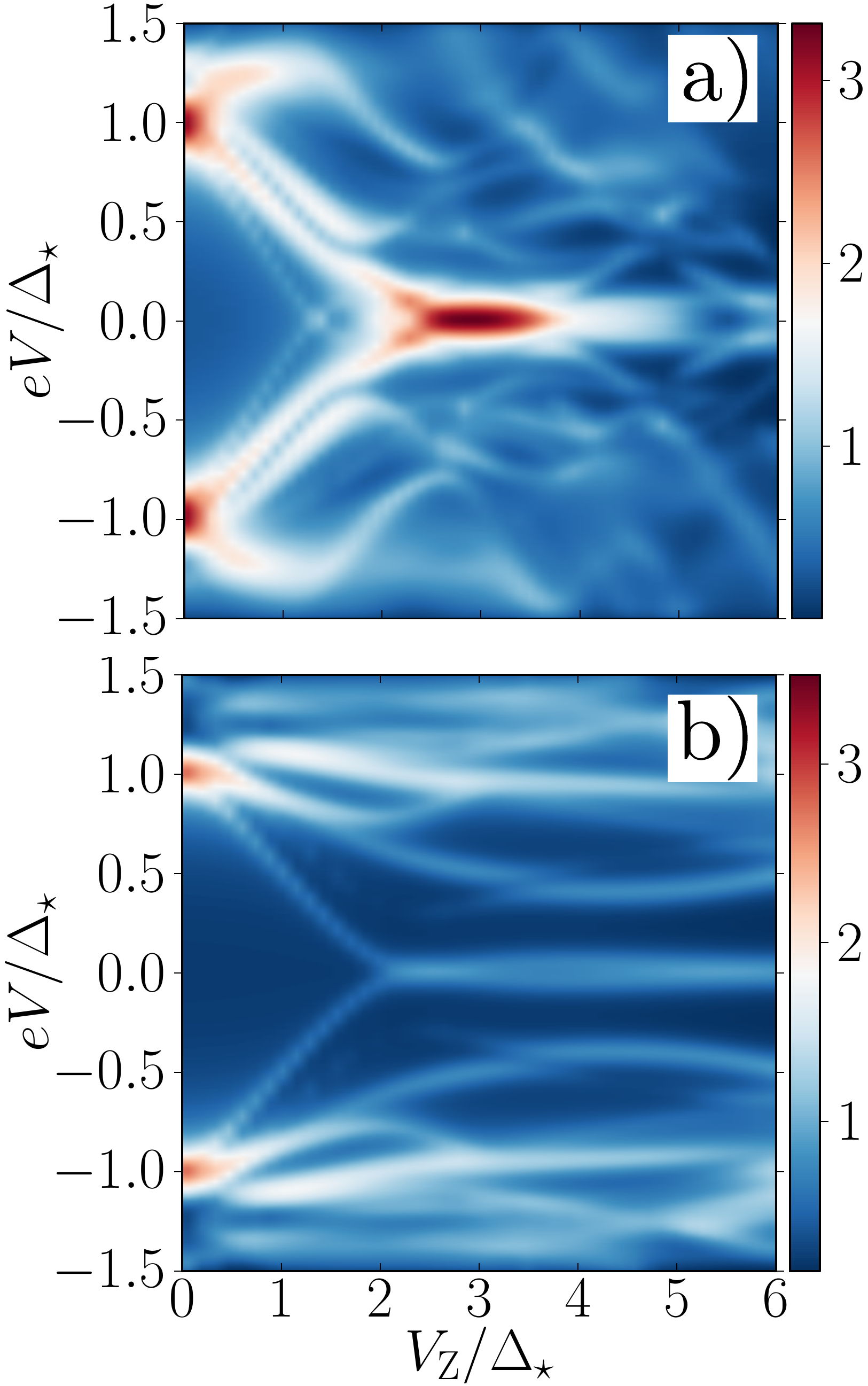}
\caption{Effect of larger SOI strength on disorder, NSS$^\prime$ case. 
The parameter values are the same as in Fig.~\ref{fig:Slow_closing_gap}. In addition, a realistic disorder $w_{\bm m}\in[-3,3]$ meV (corresponding to a mean free path $\ell_{\rm mfp}\simeq150$ nm~\cite{liu_zero-bias_2012}) is included over the entire nanowire length $L\simeq2.5~\mu$m. We do not average over disorder configurations. (a) $\alpha=0.2$ eV$\cdot$\AA. (b) $\alpha=0.8$ eV$\cdot$\AA. 
In the weak SOI regime, the disorder lowers or destroys the gap relative to lower subbands, bringing many supra-gap states down, close to the Fermi level, where they cluster in some cases into a finite-extension ZBP, like in panel (a). Such clustering is, however, removed for stronger SOI~\cite{liu_zero-bias_2012}, see panel (b).}
\label{fig:WeakSO_Disorder}
\end{figure}

(1)
Under the assumption that the measured ZBP ~\cite{mourik_signatures_2012,deng_observation_2012,das_evidence_2012} 
arises from MFs,  we conclude that $\mu\simeq0$ in the topological section,
since the ZBP emerges already at small $B$, $\frac{1}{2}g\mu_{\rm B}B\simeq\Delta_\star$ for $g=50$.

However, such a small  $\mu$, together with the reported SOI values~\cite{mourik_signatures_2012}, would generate a rapid closing of the $k_{\rm F}$-gap $\Delta_{k_{\rm F}}$ as a function of $B$. This is indeed what we find in our transport calculations for  $\mu\simeq0$, $\alpha=0.2$ eV$\cdot$\AA, both in the NS and NSS$^\prime$ setup, see Figs.~\ref{fig:Slow_closing_gap}(a) and \ref{fig:Slow_closing_gap}(c), respectively.
Note that in the NS case the ZBP stays at zero for all fields, whereas in the NSS$^\prime$ case the ZBP  exhibits an oscillating splitting (see below).
In the same figure we show that a stronger SOI gives a better agreement with the measured $\Delta_{k_{\rm F}}(B)$, both in the NS setup~\cite{prada_transport_2012}, see panel (b), and in the NSS$^\prime$ setup, shown in panel (d).  
Note that this latter SOI effect, which answers the issue raised in point (iv) above, is independent of the nature of the observed ZBP.

As already observed elsewhere~\cite{stanescu_close_2012}, the considered regime of $\mu\simeq0$ is characterized by an invisible gap closing, probably due to pretransition wave functions which are delocalized throughout the wire, with little weight close to the probed edges. At finite temperature we observe this behavior both in the NS and in the NSS$^\prime$ setups. On can thus state that issue (i) has been settled.

\begin{figure}[h!]
\includegraphics[width=0.80\columnwidth]{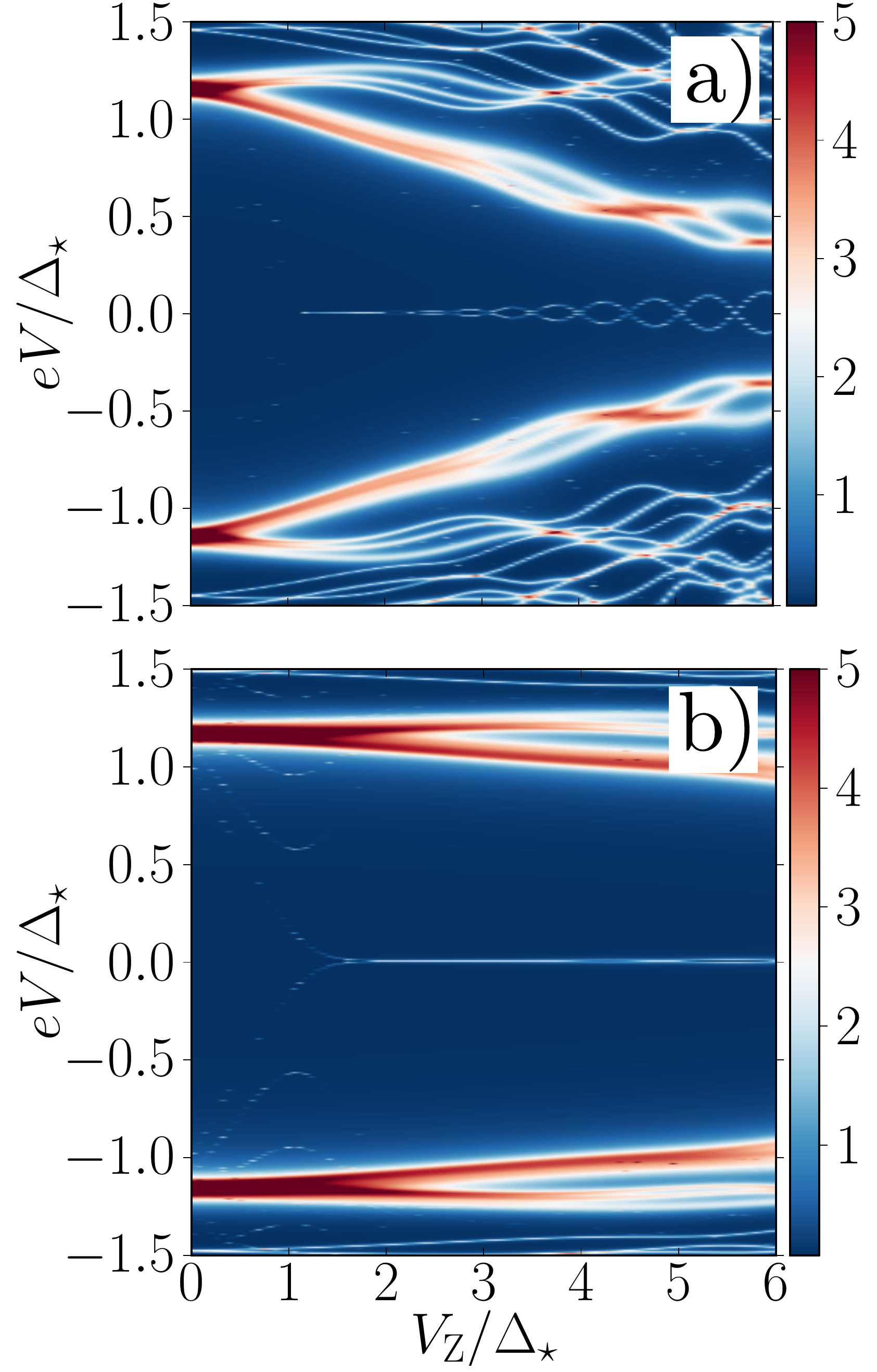}
\caption{
Same parameters as in Fig.~\ref{fig:Slow_closing_gap}, without disorder and at $T=0$. Only the wire length is slightly smaller, $L_\star=2.2\,\mu$m.
(a) $\alpha=0.2$ eV$\cdot$\AA.  Note the oscillations of the ZBP, for explanations see  text. (b) $\alpha=0.8$ eV$\cdot$\AA. For larger SOI, the oscillations become visible at higher magnetic field ($B_{\rm c}^{**}$ increases), and	
with large enough SOI strength the gap closing becomes partially visible, even for the considered case of $\mu\simeq0$, while it is not visible in an NS setup with the same parameters. The $dI/dV$ peaks coming from the gap-closing have, however, very small width and they get washed out by realistic temperatures.}
\label{fig:Closing_visible}
\end{figure}

(2)
When realistic Anderson disorder is included in the model, the closing of the gap becomes visible again even in the $\mu\simeq0$ regime~\cite{pientka_enhanced_2012,liu_zero-bias_2012,Altland_disorder_2012}, reintroducing a discrepancy with  experiments~\cite{mourik_signatures_2012,deng_observation_2012,das_evidence_2012}.
Disorder in a  nanowire with {\it weak} SOI causes a number of subgap states to appear, some of which cluster around zero energy and possibly give rise to a nontopological ZBP, more markedly for finite $\mu$~\cite{liu_zero-bias_2012}. 
Such states are coming from other subbands, for which the effective minigap gets reduced in the presence of disorder. This is substantiated by the fact that the ZBP in Fig.~\ref{fig:WeakSO_Disorder}(a) has a conductance peak larger than $2e^2/h$, implying that it cannot come from the Andreev signal of a single band.
For stronger SOI, the effect of disorder gets suppressed, and fewer subgap states are observed (see Fig.~\ref{fig:WeakSO_Disorder}), more compatibly with the experimental evidence~\cite{mourik_signatures_2012,deng_observation_2012,das_evidence_2012}. 
Due to the same mechanism, also the strong ZBP feature of Fig.~\ref{fig:WeakSO_Disorder}(a) disappears, though.
Thus, disorder is unlikely to explain the ZBP structure observed in experiments.

\begin{figure}[tb]
        \centering
        \includegraphics[width=1.0\columnwidth]{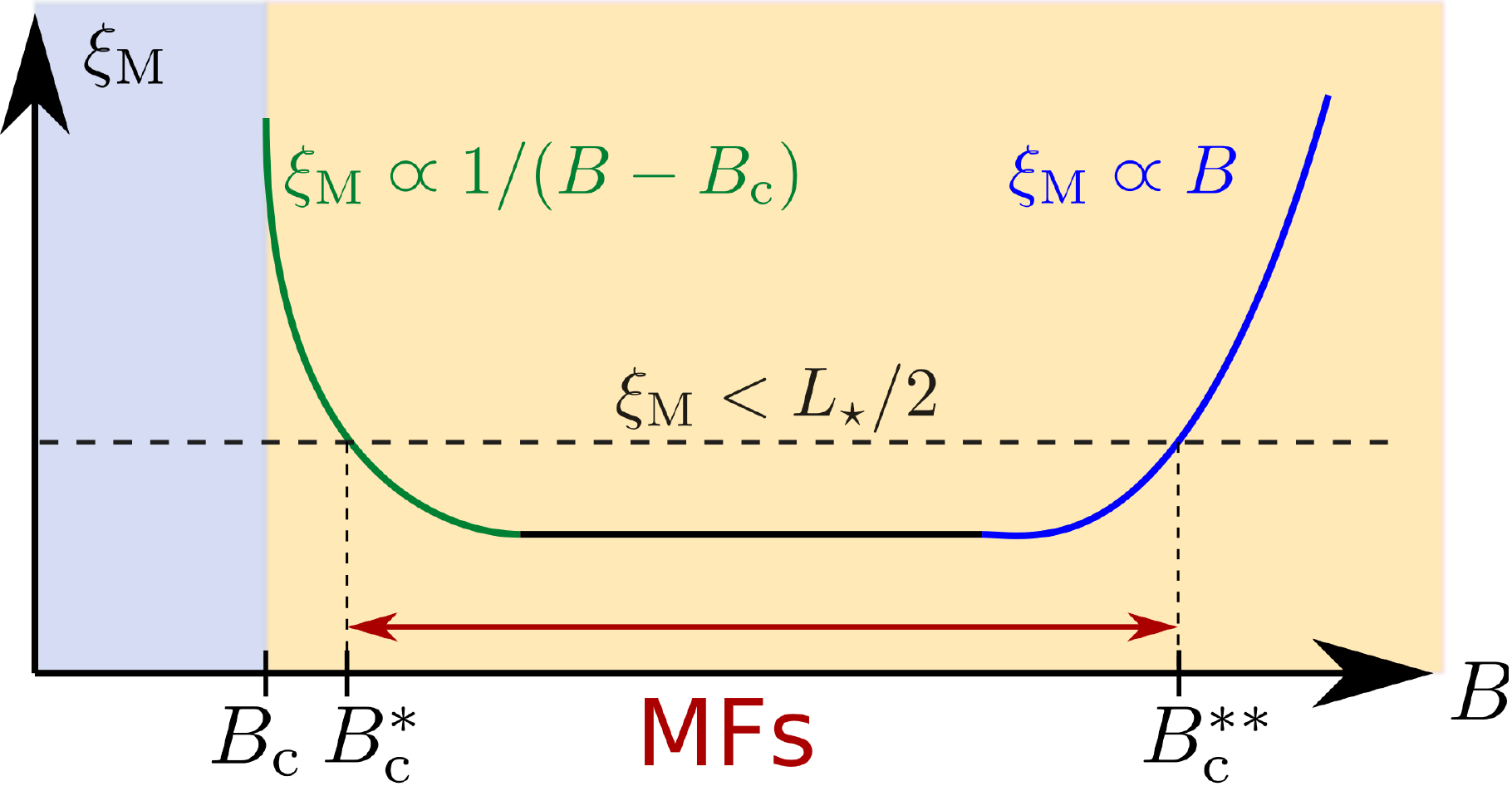}
\caption{Schematic dependence of the MF localization length
$\xi_{\rm M}$ on magnetic field $B$. According to the theory for a
one-band semi-infinite nanowire \cite{Oreg_Majorana_2010, DasSarma_Majorana_2010}, a MF emerges when the magnetic field exceeds a critical value $B_{\rm c} = 2\sqrt{\Delta_\star^2+\mu^2}/g\mu_{\rm B}$, and the system goes from the nontopological (gray) to the
topological (yellow) regime. However, for a nanowire
of finite length $L_\star$, due to overlap of the MFs from each end, the
additional approximate condition for the observation
of a MF is $\xi_{\rm M}<L_\star/2$ . Considering typical dependences of
$\xi_{\rm M}$ on magnetic field~\cite{Klinovaja_Majorana_2012}, we predict that the MF should be observed for $B_{\rm c}^*<B<B_{\rm c}^{**}$, 
where the critical fields $B_{\rm c}^*$ and $B_{\rm c}^{**}$ are defined through $\xi_{\rm M} (B_{\rm c}^*) \approx \xi_{\rm M}(B_{\rm c}^{**})\approx L_\star/2$ (cf. Fig. \ref{fig:Closing_visible}).
}
\label{Fig:localization1}
\end{figure}
\begin{figure}[h!]
\includegraphics[width=0.75\columnwidth]{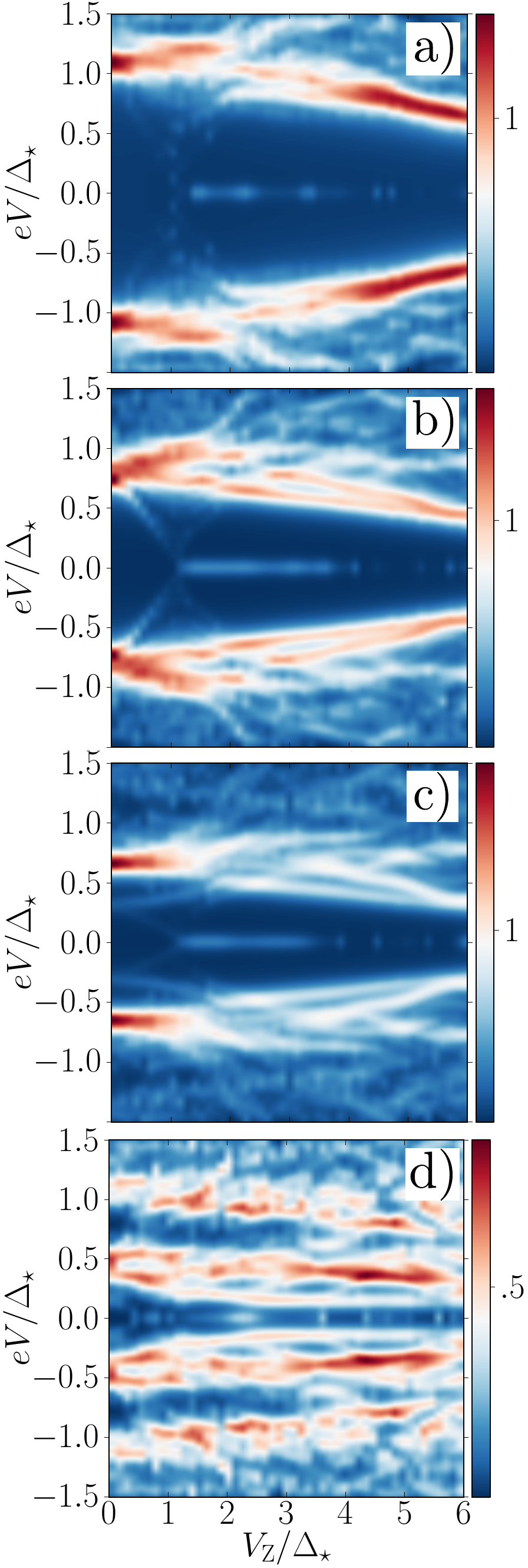}
\caption{Dependence of the conductance behavior on the normal-region length $L_{\rm N}$. Panels (a)-(d) correspond to $L_{\rm N}=0,0.3,0.6,1.5\,\mu$m, respectively. 
The other parameters are chosen as in Fig.~\ref{fig:Slow_closing_gap}, apart from: $\lambda=20$ nm, $L_\star=3~\mu$m, $\alpha=0.4$ eV$\cdot$\AA, $w_{\bm m}\in[-1.2,1.2]$ meV ($\ell_{\rm mfp}\simeq1\,\mu$m). Temperature is set to $T=75$ mK, as before. Note the evolution of the proximity peak towards lower energies and the appearance of a second peak at the largest values of $L_{\rm N}$ [panel (d)].} \label{fig:L_N}
\end{figure}

(3)  As a consequence of the finite length of the topological section ($L_\star$) and of the $B$ dependence of  $k_{\rm F}$, we observe that the ZBP splitting exhibits oscillations of increasing amplitude as $B$ is swept, see Fig.~\ref{fig:Closing_visible}(a)~\cite{oscillations}. 
To explain this, we recall that in the weak-SOI limit the MF wave function has an exponentially decaying envelope with localization length $\xi_{\rm M}$ and a fast-oscillating part $\sim \sin (k_{\rm F} x)$~\cite{Klinovaja_Majorana_2012}. 
If the magnetic field exceeds a critical value $B_{\rm c}^{**}=B_{\rm c}^{**}(\alpha,L_\star)$ (see Fig.~\ref{Fig:localization1}), the two end-MFs overlap and split away from zero energy. 
Since $\xi_{\rm M}$ increases with $B$~\cite{sato_non-abelian_2009,sau_generic_2010,sato_non-abelian_2010,alicea_majorana_2010,DasSarma_Majorana_2010,Oreg_Majorana_2010,Klinovaja_Majorana_2012}, so does the splitting.
However, if $k_{\rm F} L_\star$ becomes an integer multiple of $\pi$ as a function of $B$, the ZBP splitting returns to zero, leading to oscillations with a period given by
\begin{equation}
\delta (V_{\rm Z}/\Delta_\star)= \frac{\pi \hbar}{L_\star \Delta_\star} \sqrt{\frac{{2V_{\rm Z}}}{m }}=
\frac{\pi a}{L_\star} \frac{\sqrt{t V_{\rm Z}}}{\Delta_\star }\, ,
\end{equation}
where $m$ is the band mass and $a$ the lattice constant. 
Using parameter values corresponding to Fig.~\ref{fig:Closing_visible}, $t/\Delta_\star=40$, $L_\star/a=200$, we obtain quantitative agreement with the  simulated ZBP oscillations.
Since the critical field $B_{\rm c}^{**}$ increases with SOI~\cite{Klinovaja_Majorana_2012}, the ZBP splitting and related oscillations occur at larger fields. 
In other words, the presence or absence of the oscillations in a given range of magnetic field values is determined by the strength $\alpha$ of the SOI and by the ratio $\xi_{\rm M}(\alpha)/L_\star$. The former fixes the form of the MF wave function, the latter determines whether the two MF bound states are overlapping in a significant way or not. This explains why in Fig.~\ref{fig:Closing_visible}(b), where strong SOI has been adopted, oscillations are starting at higher $B$ (barely visible).

Note that these oscillations are quite robust against temperature effects, see Fig.~\ref{fig:Slow_closing_gap}(c).
Such  behavior of the ZBP is quite remarkable and provides an additional possible signature to identify MFs experimentally. 
To make contact with issue (ii) raised in Sec.~\ref{sec:Introduction}, one can argue at this point that the absence of oscillations in the experimentally observed ZBPs~\cite{mourik_signatures_2012,das_evidence_2012} represents an additional hint for strong SOI.

We note in passing that in the NSS$^\prime$ setup the SOI affects the visibility of the gap closing, see Fig.~\ref{fig:Closing_visible}. Again, one can explain this behavior by invoking the changing spatial profile of the wave functions close to the wire edge for different SOI values, together with the finite length of the wire.
The same effect is not manifested in the case of the NS setup (infinite wire length).\\

\begin{figure}[h!]
\includegraphics[width=0.80\columnwidth]{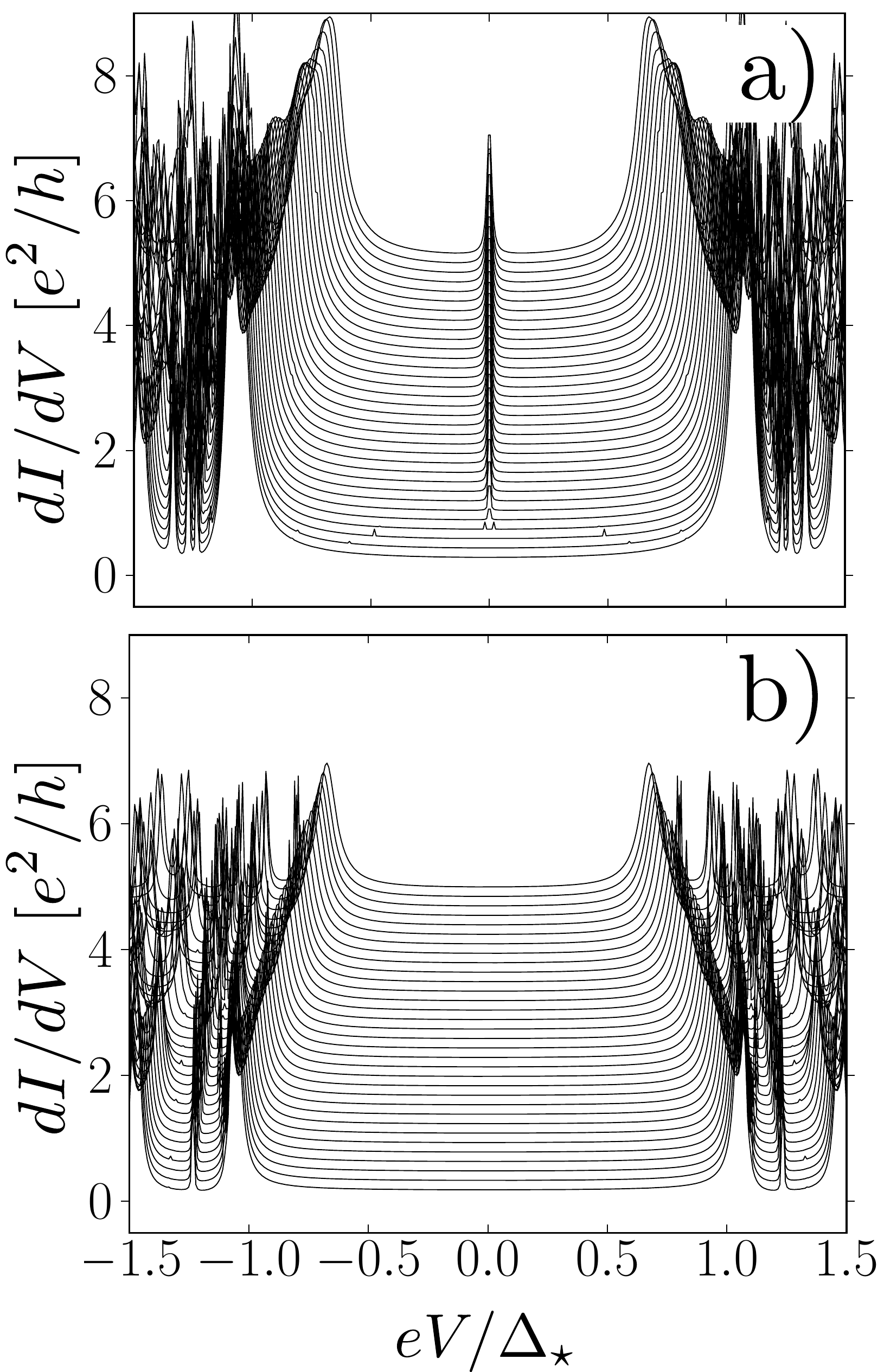}
\caption{Role of the tunnel barrier $U(x)$. The parameters are chosen as in the previous figures, different curves refer to different magnetic field values, ranging from 0 to $6\Delta_\star=1.5\,$meV ($=1\,$T for $g=50$), in steps of 0.2\,$\Delta_\star$. Here we kept $T=0$ in order to show the effect of the barrier smoothness alone. a) Gaussian tunnel barrier with width $\lambda=1$ nm (essentially, a $\delta$-function). The Majorana-induced ZBP is fully visible, with maximal weight $dI/dV=2e^2/h$ at the largest magnetic fields. The closing of the gap is, however, nearly absent. 
b) Same system, but with $\lambda=50$ nm, a value closer to the experimental situation. Gap-closing and ZBP are completely absent (the adopted energy resolution $\delta E$ is much smaller than the realistic $k_{\rm B}T$).} \label{fig:Tunnel_barrier}
\end{figure}

Next we address further issues that have received less attention in the literature so far.

(4) The position of the proximity gap $\Delta_{k_{\rm F}}(B=0)$ as observed in the $dI/dV$ curves is in general different from $\Delta_\star$ inserted
in the Hamiltonian, Eq.~(\ref{H}).  This observation is important, since it 
means that deducing  the proximity gap from the conductance curves
is not a correct procedure~\cite{mourik_signatures_2012,das_evidence_2012}.  
Such an energy shift can be due to the presence of a normal section of finite length $L_{\rm N}$ between tunnel barrier and NS interface. More precisely, the observed peak
moves to {\it lower} bias voltages for larger $L_{\rm N}$. By
increasing $L_{\rm N}$, one can move the conductance peak deeper inside the gap and eventually even introduce additional peaks when $L_{\rm N}\gtrsim \xi=\hbar v_{\rm F}/(\pi \Delta_\star)$, similarly to the case of McMillan-Rowell resonances~\cite{McMillan-Rowell}. This behavior is summarized in Fig.~\ref{fig:L_N}. 
Alternatively, in the NSS$^\prime$ configuration the peak corresponding to
$\Delta_\star$ itself can be viewed as a subgap resonance of the larger gap
$\Delta_0$, and its position can be changed by varying the distance $L_\star$ of the
N-S interface from the S-S$^\prime$ interface. In this case, the peak moves to $\it larger$ energies for decreasing $L_\star$, see, e.g., Fig.~\ref{fig:Slow_closing_gap}(c) and \ref{fig:Slow_closing_gap}(d), and Fig.~\ref{fig:Closing_visible}, where the $dI/dV$ peak is above $\Delta_\star$ due to the finite wire length [compare with Figs.~\ref{fig:Slow_closing_gap}(a) and \ref{fig:Slow_closing_gap}(b)].

(5) In both NS and NSS$^\prime$ configurations, the tunnel barrier plays an
important role --- it determines the transmission of  each transport channel,
which in turn sets the width of the subgap
resonances~\cite{McMillan-Rowell} (without changing their height).
Introduction of temperature smears the resonances while preserving their weight, which implies a reduction of the height in correspondence to the barrier-induced reduction of the width. This explains the very small value of the ZBP in experiments, and answers to issue (iii).
If the resonance width becomes smaller than the temperature, the resonance is
essentially invisible~\cite{liu_zero-bias_2012}. 
Consequently, if the barrier is wide enough, no subgap features are present in the $dI/dV$ curve. 
If the tunnel barrier is chosen to be sharp (like in many analytical and numerical calculations), all the states present in the nanowire could become visible. 
However, that is not a realistic choice, since a typical barrier in experiments has a characteristic width of $\sim50$ nm.
For such values we already observe a momentum filtering~\cite{prada_transport_2012}, leading in some cases to a complete disappearance of MF signatures, see
Fig.~\ref{fig:Tunnel_barrier}. Again, introducing disorder can make the aforementioned subgap features reappear. Therefore, it is the combined effect of barrier shape, SOI, and disorder strength that determines
the final visibility of MFs. 

\section{Conclusions}
In summary, by numerically simulating a more realistic setup than before, we have obtained new features in the transport that are similar to the ones observed in experiments. However, even after considerable effort, we do not reproduce all such features in a single configuration, and we still lack a satisfactory agreement with experiments. In particular, the exact shape of the measured ZBP is not very compatible with the picture of MFs that form and then split as a function of magnetic field.  
Thus, either the theoretical model is still incomplete, or
a different physical origin for the observed  ZBP~\cite{mourik_signatures_2012,deng_observation_2012,das_evidence_2012} is to be considered.
More precisely, from our findings it seems possible that in the experiments
the MF features are essentially invisible and the
observed ZBP is coming from some different coexisting phenomenon, like
Kondo effect, which seems indeed to yield a similar behavior in some situations~\cite{lee_zero-bias_2012}.

\section{Acknowledgments}
We thank Fabio Pedrocchi for useful discussions and Fabio Taddei for support with the numerics.
This work has been supported by the Swiss NSF, NCCR Nanoscience, NCCR QSIT,
and the EU project SOLID.


\begin{thebibliography}{30}
\bibitem{Note}
	Note that the SW region is not grounded in typical experiments, while the mean-field formalism adopted here implies this assumption. Such difference can be accounted for through a modified formula for the conductance, and leads to small corrections~\cite{LambertRaimondi1998,AnatramDatta1996}.
\end{thebibliography}
\end{document}